\documentclass[aps,prl,twocolumn,showpacs,superscriptaddress,amsmath,amssymb,longbibliography]{revtex4-1}
\usepackage[normalem]{ulem}
\usepackage{graphicx}
\usepackage{color,soul}
\usepackage{float}
\usepackage{mathtools}
\usepackage{hyperref}

\hypersetup{hidelinks}

\makeatletter 
\def\p@figure{\color{blue}} 
\def\p@equation{\color{blue}}
\def\p@table{\color{blue}}
\def\p@bibliography{\color{blue}} 
\makeatother

\graphicspath{ {./Figures/} }

\begin{document}

\title{The impact of one-way streets on the asymmetry of the shortest commuting routes}

\author{Hygor P. M. Melo} \email{hpmelo@fc.ul.pt}
\affiliation{Centro de F\'{i}sica Te\'{o}rica e Computacional, Faculdade de Ci\^encias, Universidade de Lisboa, 1749-016
Lisboa, Portugal}
\author{Diogo P. Mota}
\affiliation{Centro de F\'{i}sica Te\'{o}rica e Computacional, Faculdade de Ci\^encias, Universidade de Lisboa, 1749-016
Lisboa, Portugal}
\author{Jos\'e S. Andrade Jr.}
\affiliation{Departamento de F\'isica, Universidade Federal do Cear\'a, 60451-970, Fortaleza, Cear\'a, Brazil}
\author{Nuno A. M. Ara\'ujo} \email{nmararaujo@fc.ul.pt}
\affiliation{Centro de F\'{i}sica Te\'{o}rica e Computacional, Faculdade de Ci\^encias, Universidade de Lisboa, 1749-016
Lisboa, Portugal}
\affiliation{Departamento de F\'isica, Faculdade de Ci\^encias, Universidade de Lisboa, 1749-016
 Lisboa, Portugal}


\begin{abstract}
On a daily commute, the shortest route from home to work rarely overlaps completely the shortest way back. 
We analyze this asymmetry for several cities and show that it exists even without traffic, due to a non-negligible fraction of one-way streets. For different pairs of origin-destination ($\rm OD$), we compute the log-ratio $r=\ln(\ell_{\rm D}/\ell_{\rm O})$, where $\ell_{\rm O}$ and $\ell_{\rm D}$ are the lengths of the shortest routes from $\rm O$ to $\rm D$ and from $\rm D$ to $\rm O$, respectively. While its average is zero, the amplitude of the fluctuations decays as a power law of the $\rm OD$ shortest path length, $r\sim \ell_{\rm O}^{-\beta}$. Similarly, the fraction of one-way streets in a shortest route also decays as $\ell_{\rm O}^{-\alpha}$. Based on semi-analytic arguments, we show that $\beta=(1+\alpha)/2$. Thus, the value of the exponent $\beta$ is related to correlations in the structure of the underlying street network.
\end{abstract}

\maketitle

\section{Introduction}

A large fraction of our day is wasted traveling between home and work. On average, European and US workers spend more than 25 minutes each way~\cite{EuroStat,burd2021travel}, a value that is increasing every year, with a major impact in the local economy and public health~\cite{lyons2008human,pereira2015commute}. The usual strategy to minimize this impact is obviously to choose the shortest route. However, the shortest route from home to work does not always overlap completely  the way back. This asymmetry stems from a non-negligible fraction of one-way streets. One-way streets are designed for safety reasons and to mitigate heavy traffic~\cite{venerandi2017form,stemley1998one}. They affect the length distribution of optimal routes~\cite{verbavatz2021one} and have a critical impact on the resilience of street networks to traffic jams~\cite{carmona2020cracking}.

Here, we study the asymmetry between the length of the two shortest routes. To characterize this asymmetry, we defined a log-ratio $r$, 
\begin{equation}
r=\ln \left(\frac{\ell_{\rm D}}{\ell_{\rm O}}\right),
\label{equation_1}
\end{equation}  
where $\ell_{\rm O}$ is the length of the shortest route from the origin to the destination, and $\ell_{\rm D}$ is the length of the shortest route on the way back. We analyze ten cities worldwide, with more than one million inhabitants in their metropolitan area, and find that the average of $r$ vanishes for all values of $\ell_{\rm O}$, while its standard deviation $\sigma_r$ behaves as a decreasing function of $\ell_{\rm O}$, consistent with a power-law decay $\ell_{\rm O}^{-\beta}$. Moreover, in order to evaluate the impact of one-way streets, we measure the average fraction of the length corresponding to one-way streets in each shortest route, $\langle f \rangle_{\ell_{\rm O}}$. We find that this fraction also decays algebrically with the total path length $\ell_{\rm O}$, $\langle f \rangle_{\ell_{\rm O}} \sim \ell_{\rm O}^{-\alpha}$. By assuming that the differences in the shortest routes are only due to one-way streets, we derive the scaling relation $\beta=(1+\alpha)/2$ and show that it is consistent with the empirical data.

\begin{figure}
\includegraphics[width=8.0cm]{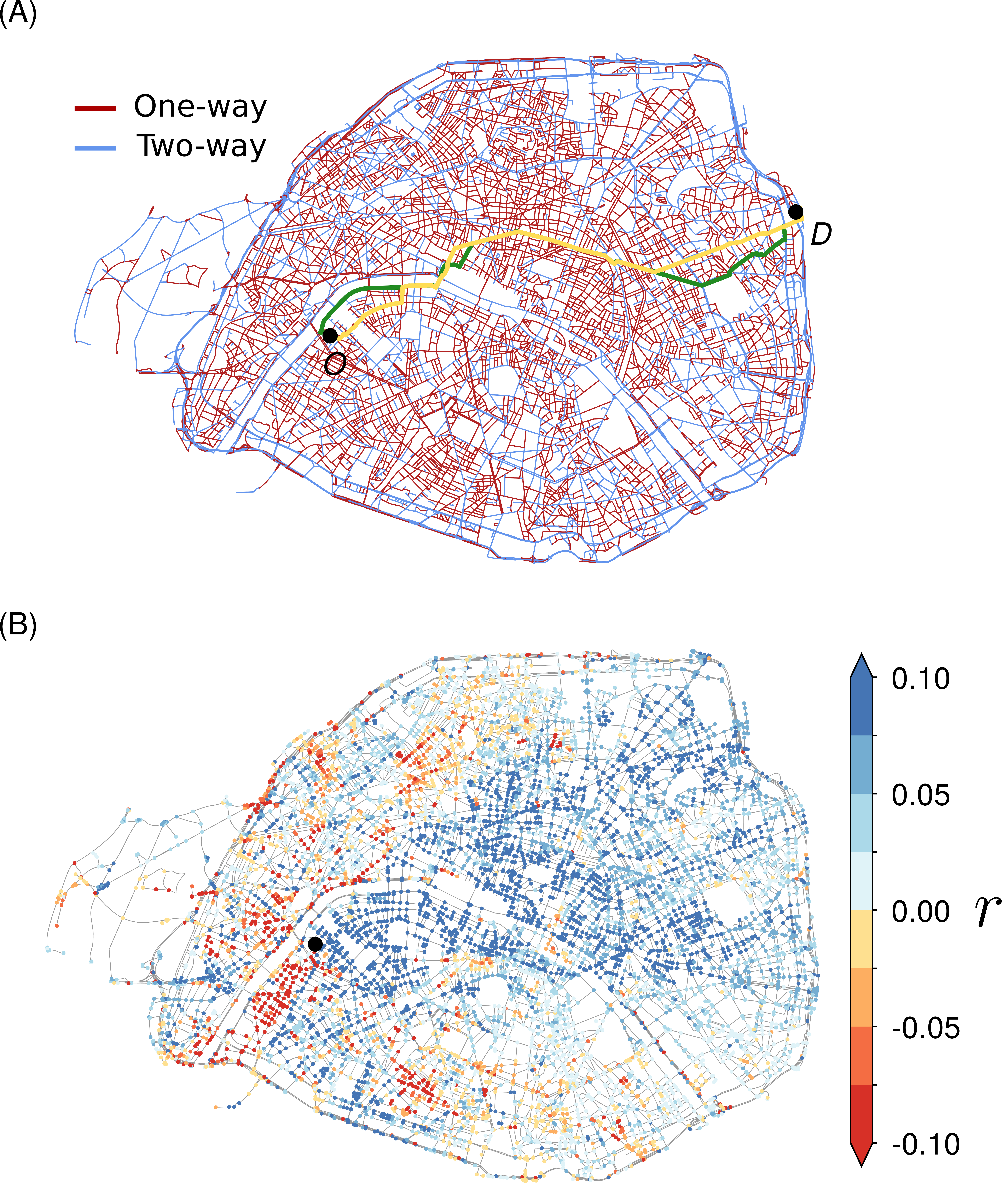}
\caption{\textbf{One-way streets and asymmetry in the shortest routes.} Street networks can be mapped into graphs, where the edges are road segments and vertices are the intersection of those segments. The edges are classified as one way (directed edge) or two way (undirected edge). (A) Street map of Paris where one-way segments are represented in red and two-way segments are in blue. The yellow line is the shortest route from origin $\rm O$ to destination $\rm D$, with a length $\ell_{\rm O}$, and the green line is the shortest way back, with a length $\ell_{\rm D}$. This asymmetry is a consequence of the existence of one-way streets, and can be quantified in terms of the log-ratio $r=\ln\left(\ell_{\rm D}/\ell_{\rm O}\right)$. (B) Street map of Paris where the color is the value of $r$ from an origin marked by the black circle. The distribution of $r$ is clearly heterogeneous and for some destinations differences in lengths can be greater than 10\%.}
\label{fig_1}
\end{figure}
\begin{figure*}[t]
\includegraphics[width=18.0cm]{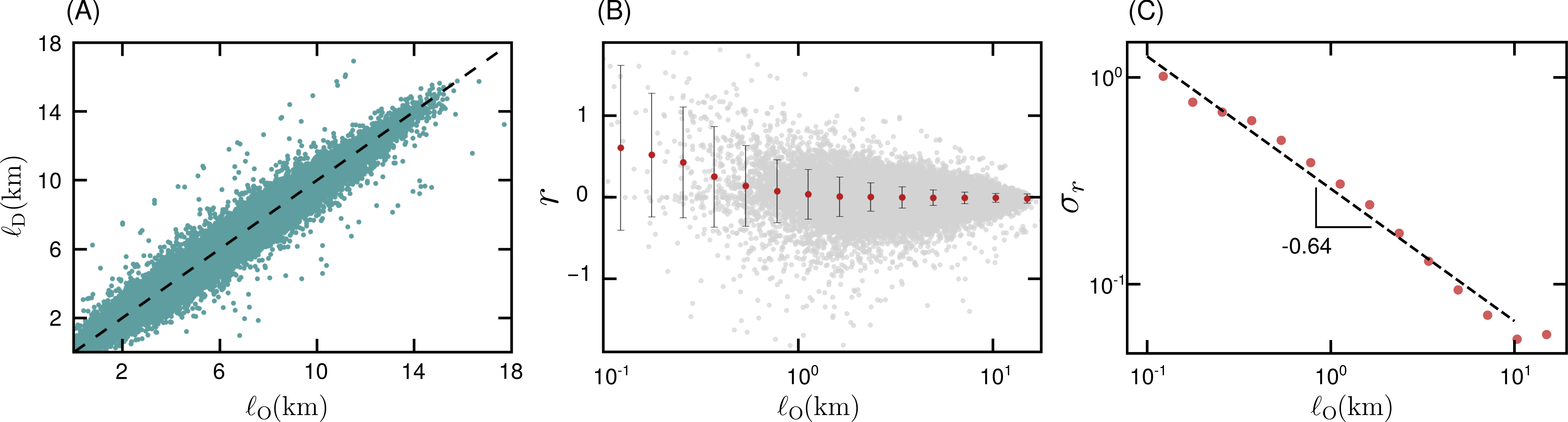}
\caption{\textbf{Fluctuations of $r$ for the city of Paris.} For each origin-destination pair,
we have the shortest route length $\ell_{\rm O}$ from the origin to the destination and the shortest route length $\ell_{\rm D}$ from the destination to the origin. In (A) we see the relation between $\ell_{\rm D}$ and $\ell_{\rm O}$ for $10^5$ random $OD$ pairs in Paris. The data points are concentrated along the (dashed) symmetric line, where $\ell_{\rm D}=\ell_{\rm O}$, but with non-negligible fluctuations around it. To quantify the fluctuations, we first define the logarithm ratio $r=\ln(\ell_{\rm D}/\ell_{\rm O})$ and calculate the standard deviation $\sigma_r$ for a set of log-spaced bins on $\ell_{\rm O}$. (B) The average of $r$ (red circles) converges to zero within about 1 km. However, the fluctuations $\sigma_r$, indicated by the bars, tend to decrease smoothly with $\ell_{\rm O}$. (C) $\sigma_r$ decays as a power law, $\sigma_r\sim \ell_{\rm O}^{-\beta}$, with $\beta=0.64\pm 0.04$. The dashed line represents the best power-law fit. The exponent was calculated by averaging over the least-squares fit of $100$ bootstrapping samples. The error is one standard deviation.}
\label{fig_2}
\end{figure*}
\begin{figure}[t]
\includegraphics[width=7.0cm]{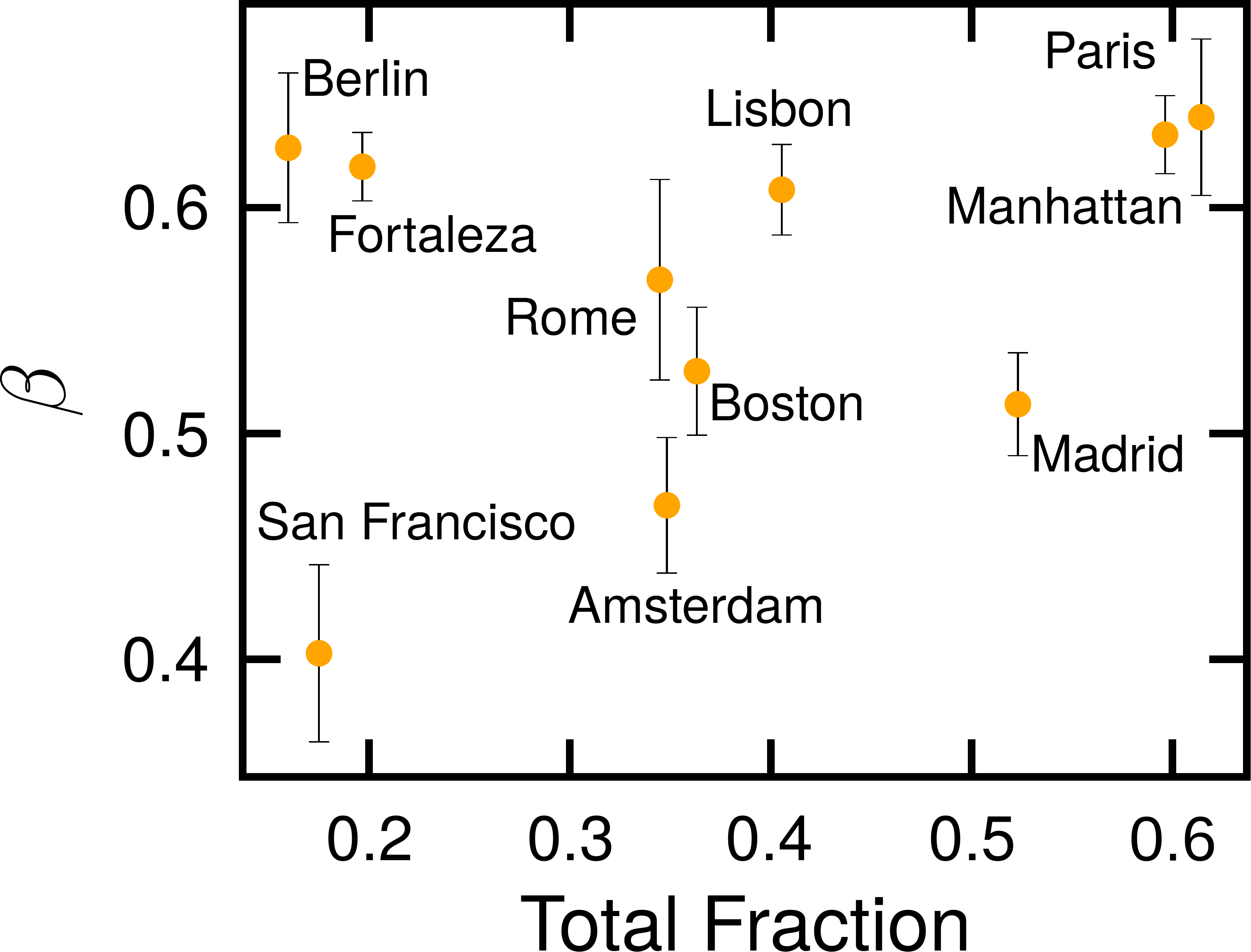}
\caption{\textbf{Exponent $\beta$ and total fraction of one-way streets.} 
The exponent $\beta$ as a function of the total length fraction of one-way streets for each city. We see no clear relationship between $\beta$ and the total fraction of one-way streets, a global property of the city. The correlation is weak, with a Person coefficient of 0.32.}
\label{fig_3}
\end{figure}
\begin{figure*}[t]
\includegraphics[width=17.5cm]{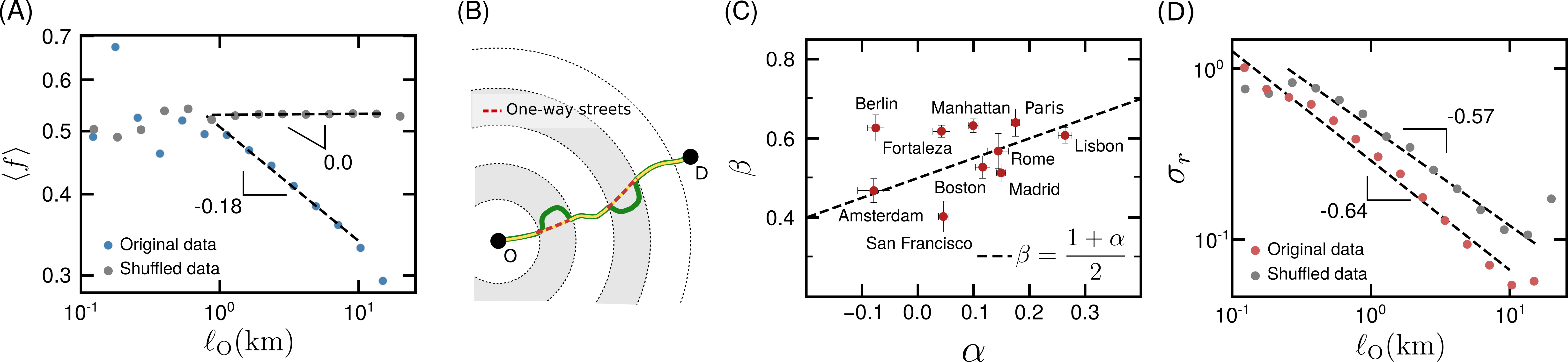}
\caption{\textbf{Scaling and the fraction of one-way streets.} 
To explain the values of $\beta$ we look at the local distribution of one-way segments along the shortest routes. In (A) we show that $\langle f \rangle_{\ell_{\rm O}}$ along the shortest route decays as a power law $\langle f\rangle_{\ell_{\rm O}}\sim \ell_{\rm O}^{-\alpha}$, with an exponent $\alpha=0.18\pm 0.01$ for Paris (blue circles).  The dashed line represents the best power-law fit. The same is true for the other cities. The nonuniform distribution of one-way segments along the shortest routes determines the amplitude of the fluctuations $\sigma_r$, fixing a relationship between the exponents $\beta$ and $\alpha$. In (B) we show a schematic representation of the effect of the distribution of one-way segments and the fluctuations on $\ell_{\rm D}$. Given a pair origin-destination ($\rm OD$), there is a shortest route in yellow from $\rm O$ to $\rm D$ with length $\ell_{\rm O}$, and a shortest route in green from $\rm D$ to $\rm O$ with length $\ell_{\rm D}$. From the definition of $r$, we show that the amplitude of fluctuations $\sigma_r$ is bounded by the fluctuations on $\ell_{\rm D}$, in the form of $\sigma_r=\sigma_{\ell_{\rm D}}/\ell_{\rm O}$. To determine $\sigma_{\ell_{\rm D}}$, we assume that there is a divergence of the two routes on the region where there are one-way segments along $\rm O$ to $\rm D$. Dividing the route in equal parts showed that the fluctuations on $\ell_{\rm D}$ scale as $\sigma^2_{\ell_{\rm D}}\sim \langle f \rangle_{\ell_{\rm O}} \ell_{\rm O}$, where $\langle f \rangle_{\ell_{\rm O}}$ is the average fraction of one-way segments on routes of length $\ell_{\rm O}$. In (C) the  exponent $\beta$ is plotted against the exponent $\alpha$ for different cities. The black dashed line, which  corresponds to the relation $\beta=(1+\alpha)/2$, shows good agreement with the data. The presence and importance of spatial correlations can be measured by shuffling the position of one-way streets in the road network.  We see in (D) that, for Paris, the exponent $\beta$ of the fluctuations in $r$ changes from $0.64$ to a value more close to $0.5$ (gray circles). Precisely, from the least-squares fit to the data, we obtain an exponent of $\beta=0.57\pm 0.02$. In order to confirm consistency with the scaling relationship $\beta=(1+\alpha)/2$, we show in (A) that the fraction of one-way segments for the shuffled map is uncorrelated with $\ell_{\rm O}$ (gray circles), resulting in an exponent for Paris $\alpha=0.001\pm 0.002$. The exponent was calculated by averaging over the least-squares fit of $100$ bootstrapping samples. The error is one standard deviation.}
\label{fig_4}
\end{figure*}

\section{Results}

Street networks are directed graphs, where the edges are road segments that are classified as one-way (directed) or two-way (undirected) segments (edges). The vertices are the intersection between those segments. Using the python library OSMnx~\cite{boeing2017osmnx}, we collected from the OpenStreetMap~\cite{OpenStreetMap} the street network of ten cities worldwide including their neighbors, as defined by the administrative borders. From the OpenStreetMap, each edge has a set of attributes such as the type of road, length, and if the road is of one or two ways. For some cases ({\it e.g.}, avenues), two parallel and close segments are labeled in the original data as two one-way streets, but we consider it here as a single two-way segment (see Materials and Methods). As an example, we show in Fig.~\ref{fig_1}(A) the spatial distribution of one and two-way segments in Paris.

Also shown in Fig.~1(A) is an example of the shortest path from an origin $\rm O$ and a destination $\rm D$ (yellow line) in Paris and the corresponding way back (green line). Clearly, one sees that the two do not overlap for some parts of the journey. As already mentioned, here we choose to locally quantify this asymmetry using the log-ratio $r$ defined in Eq.~(\ref{equation_1}). When $\ell_{\rm D}\approx \ell_{\rm O}$, $r\approx\frac{\ell_{\rm D}-\ell_{\rm O}}{\ell_{\rm O}}$, and therefore, $r$ measures the relative difference between $\ell_{\rm D}$ and $\ell_{\rm O}$. From a specific origin (black circle in Fig.~\ref{fig_1}(B)), we estimate $r$ for all vertices, where the color of the vertices in the figure is given by the corresponding value of $r$. For certain regions, the length of the shortest route to a destination can differ by more than 10\% from its way back.

To study the distribution of $r$ in the entire city, we randomly selected $10^5$ pairs of points on each city and calculate $\ell_{\rm O}$ and $\ell_{\rm D}$. The relationship between them for Paris is shown in Fig.~\ref{fig_2}(A). We see that $\ell_{\rm D}$ and $\ell_{\rm O}$ are positively correlated. Notwithstanding,   
 we observe fluctuations around the symmetric line $\ell_{\rm D}=\ell_{\rm O}$. The log-ratio $r$ quantifies deviations from the symmetric line $\ell_{\rm D}=\ell_{\rm O}$ for each pair of origin-destination. In Fig.~\ref{fig_2}(B) we see $r$ as a function of $\ell_{\rm O}$ for Paris.
We observe that the average of $\left \langle r\right \rangle$ converges to zero within about 1 km, but the standard deviation $\sigma_r=\sqrt{\left \langle r\right \rangle^2-\left \langle r^2\right \rangle}$, represented by the error bars, decays more smoothly with $\ell_{\rm O}$. Figure~\ref{fig_2}(C) shows that $\sigma_r$ is consistent with a power-law decay, 
\begin{equation}
\sigma_r\sim \ell_{\rm O}^{-\beta},
\label{equation_2}
\end{equation}
with $\beta=0.64\pm 0.04$.
The exponent was estimated from averaging the exponent obtained for $100$ bootstrapping samples of a least-squares fit and the error bar corresponds to one standard deviation.

We measured the fluctuations $\sigma_r$ for all cities and estimated that the values of the exponent $\beta$ ranges from $0.40\pm 0.04$ to $0.64\pm 0.04$ (see Supplemental Material). To evaluate the relation with the distribution of one-way streets, we first compare the value of $\beta$ with the corresponding fraction of one-way streets in the city. As shown in Fig.~\ref{fig_3}, there is no clear relation between the two. The correlation is rather weak, with a Person coefficient of $0.32$.

The shortest route is an optimal route connecting an origin $\rm O$ and a destination $\rm D$. The fraction of one-way streets in the shortest route is not necessarily the same as the one in the entire network. In fact, our analysis reveals that this fraction is a decaying function of the length of the shortest route, as shown in Fig.~\ref{fig_4}(A). The longer the route is, the lower is the fraction of the path that goes through one-way streets. This is consistent with the fact that main roads are usually two-way streets~\cite{gayah2012analytical, stemley1998one, walker2000downtown}. The data suggests a power-law decay, $\left \langle f \right \rangle_{\ell_{\rm O}}\sim \ell_{\rm O}^{-\alpha}$, where $\alpha$ is an exponent that depends on the city. Figure~\ref{fig_4}(A) shows the average fraction of one-way segments along the optimal routes as a function of $\ell_{\rm O}$ for Paris, and the calculated exponent $\alpha=0.18 \pm 0.01$.

To establish a quantitative relation between $\beta$ and $\alpha$ we need to connect $\sigma_r$ and the fluctuations 
of $\ell_{\rm D}$ defined by $\sigma_{\ell_{\rm D}}= \sqrt{\left \langle \ell_{\rm D}\right \rangle^2-\left \langle \ell_{\rm D}^2\right \rangle}$. 
Under the approximation of small differences in the shortest routes ($\ell_{\rm O}\approx \ell_{\rm D}$), we have that 
\begin{equation}
\frac{\ell_{\rm D}}{\ell_{\rm O}}=e^r\approx 1+r.
\label{equation_5}
\end{equation}
Therefore, the length of the return route is $\ell_{\rm D}= \ell_{\rm O}+r\ell_{\rm O}$. For a fixed $\ell_{\rm O}$, $\left \langle \ell_{\rm D}\right \rangle=\ell_{\rm O}+\ell_{\rm O}\left \langle r\right \rangle$ and $\left \langle \ell_{\rm D}^2\right \rangle=\ell_{\rm O}^2+2\ell_{\rm O}^2\left \langle r\right \rangle+\ell_{\rm O}^2\left \langle r^2\right \rangle$.
Combining the two, we obtain that the standard deviation $\sigma_r$ of the log-ratio is
\begin{equation}
\sigma_{r}=\frac{\sigma_{\ell_{\rm D}}}{\ell_{\rm O}}.
\label{equation_6}
\end{equation}

In that limit, we can estimate how $\sigma_{\ell_{\rm D}}$ scales with ${\ell_{\rm O}}$ in the following way.
As schematically shown in Fig.~\ref{fig_4}(B), we divide the route in $N$ equal parts, where $N$ is proportional to $\ell_{\rm O}$, and the shortest way back will be different every time it crosses a region with one-way streets. Such events occur $N \left \langle f\right \rangle_{\ell_{\rm O}}$ times over the path. Therefore, 
\begin{equation}
\ell_{\rm D}=\ell_{\rm O}+\sum_i^{N \left \langle f\right \rangle_{\ell_{\rm O}}} \eta_i,
\label{equation_7}
\end{equation}
where $\eta_i$ are small differences in length between the trajectory on the two shortest routs. As a result, the standard deviation $\sigma_{\ell_{\rm D}}$ is 
\begin{equation}
\sigma_{\ell_{\rm D}}^2=\sum_{i,j}^{N \left \langle f\right \rangle_{\ell_{\rm O}}} \left \langle (\eta_i-\overline{\eta})(\eta_j-\overline{\eta})\right \rangle,
\label{equation_8}
\end{equation}
where $\overline{\eta}$ is the average of $\eta$.

Assuming that $\eta_i$ is a random uncorrelated variable, then $\left \langle (\eta_i-\overline{\eta})(\eta_j-\overline{\eta})\right \rangle = \Gamma \delta_{ij}$, where $\delta_{ij}$ is the Kronecker delta and $\Gamma$ is a constant that sets the amplitude of the fluctuations. 
Thus, asymptotically 
\begin{equation}
\sigma_{\ell_{\rm D}}^2\sim N\left \langle f\right \rangle_{\ell_{\rm O}}\sim \ell_{\rm O} \left \langle f\right \rangle_{\ell_{\rm O}}.
\label{equation_9}
\end{equation}
Since $\left \langle f\right \rangle_{\ell_{\rm O}}\sim \ell_{\rm O}^{-\alpha}$, and using Eq.~(\ref{equation_6}) and Eq.~(\ref{equation_9}), it follows that 
\begin{equation}
\sigma_r\sim{\ell_{\rm O}^{-(1+\alpha)/2}}.
\label{equation_10}
\end{equation} 
Comparing with Eq.~(\ref{equation_2}), we obtain the scaling relation $\beta=(1+\alpha)/2$.

Figure~\ref{fig_4}(C) demonstrates that for all cities the exponents
$\beta$ and $\alpha$ are in good agreement with the scaling relation indicated by the dashed line. 
The largest deviations are for Berlin, Fortaleza and San Francisco. These three cities are the ones with the lowest fractions of one-way streets (see Fig.~\ref{fig_3}),
with less statistics to estimate $\langle f \rangle_{\ell_{\rm O}}$ and, therefore, the exponent $\alpha$.

For the case of a random distribution of one-way segments along the shortest route, we expect  
$\alpha=0$. In this limit, $\beta=0.5$. Thus, differences in the value of the exponents $\alpha$ and $\beta$ could be due to underlying spatial correlations in the system. To test this hypothesis, we shuffled the position of one-way streets (see Materials and Methods) and repeated the analysis for $\sigma_r$. Figures~\ref{fig_4}(A) and (D) shows that, for the shuffled case, $\beta=0.57\pm 0.02$ and $\alpha=0.001\pm0.002$, suggesting that the observed exponents are due to spatial correlations in the distribution of one-way streets. The same is observed to all other cities as shown in the Supplemental Material.


\section{Conclusion}

Mobility is central for understanding the modern urban economy~\cite{di2018sequences, gonzalez2008understanding, schneider2013unravelling}. The level of mobility can be a proxy for different social aspects such as health~\cite{myers2021measuring}, criminal activity~\cite{caminha2017human}, lifestyle~\cite{di2018sequences}. Furthermore, differences on the level of accessibility can increase social and economical inequalities~\cite{moro2021mobility}. All these aspects are deeply connected to the structure of the underlying street network. The global structures and patterns of street networks are well reported~\cite{barthelemy2011spatial,strano2012elementary,boeing2018measuring,boeing2019urban,boeing2021spatial}, and there is a growth of interest in understanding the impact of one-way streets on mobility~\cite{carmona2020cracking,verbavatz2021one}. Here, we investigated the asymmetry of shortest routes originated from one-way streets. We show that the shortest route from a given origin to a destination is not necessarily the same as the way back. Although on average this asymmetry is negligible, its fluctuations are not. As a matter of fact, they decay as a power-law of the origin-destination shortest route length with an exponent $\beta$ that is related to the distribution of one-way streets on the shortest route.

Scale invariance is a general feature found in the analysis of several data for cities, such as population growth, area, and mobility~\cite{rozenfeld2008laws, makse1995modelling, song2010modelling, rozenfeld2011area, small2011spatial, louf2014congestion, barthelemy2016structure}. 
For the growth of cities, it is reported that fluctuations in the logarithm of the growth rate, defined as $\ln(S_1/S_0)$, where $S_0$ and $S_1$ are the initial and final city population, also decreases as a power law of the city size. However, due to the long-range spatial correlations in the population, the exponent is lower than $0.5$~\cite{rozenfeld2008laws}. 
Somehow different, here we also find values of $\beta$ above this value. This is likely because here we have an optimization process where the origin and destination are fixed. Thus, fluctuations occur under an additional constraint that is not present in the case of cities where the fluctuations are only limited by spatial restrictions in the population density ~\cite{rozenfeld2008laws}.


To explain $\beta$, we looked to the distribution of one-way streets within the shortest routes and found that the average fraction of one-way streets in a shortest route also scales as a power law of the route length, with an exponent $\alpha$. We demonstrated that there is a scaling relation between these two exponents $\beta=(1+\alpha)/2$, in good agreement with the street network data. This scaling relation connects the fluctuations in the shortest route length with the distribution of one-way streets, shedding light on possible large scale effects of one-way streets on mobility.

\section{Materials and Methods}

\subsection{Network processing and shortest routes}

The street network data was obtained using OSMnx~\cite{boeing2017osmnx}. OSMnx is a free, open source python package that downloads administrative boundary geometries, building footprints and street networks from OpenStreetMap. By specifying a set of queries, such as name of the city or network type, we downloaded the street networks. In this paper, we studied the following cities: Amsterdam, Berlin, Boston, Fortaleza, Lisbon, Madrid, Manhattan, Paris, Rome, and San Francisco. The network data comprises vertices and edges. The largest city in the study is Berlin with a total of $27979$ vertices and $72956$ edges, whereas Paris is the smallest with $10016$ vertices and $19487$ edges. All the analyses performed required data from a set of shortest routes between pairs of vertices. We selected $10^5$ pairs of origins and destinations vertices at random and, using Dijkstra's algorithm from NetworkX~\cite{hagberg2008exploring}, calculated two shortest routes for every pair of vertices: the origin-destination (whose length is $\ell_{\rm 0}$) and the destination-origin route (whose length is $\ell_{\rm D}$). In order to avoid any bias created by the finite size of each street network, we included in all networks the neighbors cities defined by each administrative border. All pairs of points are chosen inside the main city, but the routes can go outside the boundaries of the city. Finally, using the length of these routes, we calculate $r$ (see Eq.~(\ref{equation_1})) as a measure of symmetry between the two shortest routes.

\subsection{Defining one-way streets and shuffling}

The edges have a given set of attributes, such as the type of edge (motorway, trunk, primary, secondary, tertiary, unclassified, and residential), the beginning and end vertices, the length, and  whether an edge is one way or two way. Since every edge has a beginning vertex $u$, an end vertex $v$, and a key $k$ (to distinguish between parallel edges that connect the same $(u,v)$ pair), an edge is classified in the downloaded data as being two way if and only if there exists an $(u,v,k)$ and its inverse $(v,u,k)$ with the same length. For the case of avenues, we see that the pair segments (going in opposite ways) that constitute a part of avenue do not connect the same pair of vertices, therefore the resulting default classification from the OpenStreetMap of these edges is one way. However, the same length is traveled when moving along an avenue going from $\rm O$ to $\rm D$ and $\rm D$ to $\rm O$, meaning that from the point of view of our algorithm of navigation these edges are considered to be two ways. We reclassified the edges according to the following criteria. Edges with a lane number higher or equal to three and whose highway was classified as ``unclassified'' were reclassified as two-way edges. Additionally, any edge with the highway attribute of ``trunk'', ``primary'' or ``motorway'' was automatically reclassified as a two-way edge. To remove the correlations on the one-way street positions, we shuffled the one-way and two-way edge attributes. For each city, we used ten shuffled street networks with $10^4$ pairs of origins and destinations vertices at random. 


\begin{acknowledgments}
We acknowledge financial support from the Portuguese Foundation for Science and Technology (FCT) under Contracts No. PTDC/FIS-MAC/28146/2017 (LISBOA–01–0145–FEDER–028146), UIDB/00618/2020, and UIDP/00618/2020. JSA acknowledges the Brazilian agencies CNPq, CAPES and FUNCAP, and the National Institute of Science and Technology for Complex Systems (INCT-SC) in Brazil for financial support.
\end{acknowledgments}

\bibliography{referencias}

\end{document}